\documentclass{iopart}

\usepackage{graphicx}
\usepackage{amssymb}

\begin{document}

\newcommand \be {\begin{equation}}
\newcommand \ee {\end{equation}}
\newcommand \bea {\begin{eqnarray}}
\newcommand \eea {\end{eqnarray}}

\letter{An exactly solvable dissipative transport model}

\author{Eric Bertin}

\address{Department of Theoretical Physics,
University of Geneva, CH-1211 Geneva 4, Switzerland}
\begin{abstract}
We introduce a class of one-dimensional lattice models in which a quantity,
that may be thought of as an energy, is either transported from one site
to a neighbouring one, or locally dissipated. Transport is controlled by a
continuous bias parameter $q$, which allows us to study symmetric as well as
asymmetric cases. We derive sufficient conditions for the factorization
of the $N$-body stationary distribution and give an explicit solution
for the latter, before briefly discussing physically relevant situations.
\end{abstract}

\section*{Introduction}

Systems driven in a non-equilibrium steady state by an external forcing
that generates an internal flux are very frequent in nature.
Among these systems, two broad classes may be distinguished according to their
conservation properties. The first class corresponds to systems in which the
flux of a conserved quantity (e.g., particles or mass) takes place.
Such systems have raised considerable interest recently, and different
paradigmatic models for which the full ($N$-body) stationary probability
distribution can be worked out exactly have emerged from these studies:
the Asymmetric Simple Exclusion Process (ASEP)
\cite{Spitzer,Derrida,Sandow,Liggett,Schutz},
the Zero Range Process (ZRP) \cite{Spitzer,Evans-Rev00,Evans-Rev05},
the Asymmetric Random Average Process (ARAP)
\cite{Krug,Rajesh,Coppersmith,Zielen}, as well as more general
mass transport models \cite{Evans04b,Zia04}.

On the other side, a second class corresponds to situations
where the quantity moving through the system is locally non conserved
in the bulk.
This may happen for instance when the system can exchange particles
with a reservoir, or when one considers an energy flux in a system
where the `microscopic' dynamics --at the chosen level of description--
is already dissipative, like in turbulent flows or in shaken granular
materials for instance.
Within this class of systems, attention has been mainly devoted to
ASEP or ZRP models where particles can be added or removed within the bulk,
at rates that may differ from the boundary ones
\cite{Evans-Rev05,Evans02,Evans03,Evans04a,Levine04,Angel05}.
Yet, there seems to be very few known solvable models, for which the
probability of any microscopic configuration can be computed exactly,
where a continuous quantity is injected at the boundaries and (partially)
dissipated in the bulk, a situation of broad physical interest
\cite{Farago}
\footnote{Note that a `reverse' situation, where a
quantity is constantly added in the bulk,
has been studied in the context of force fluctuations in beads
packs, the vertical axis playing the role of time \cite{Coppersmith}.}.

In this note, we consider a generalisation of the one-dimensional
cascade model introduced in \cite{Bertin}, where a local quantity,
injected at the boundaries, is either transported or dissipated
in the bulk.
A bias parameter $q$ allows us to consider partially asymmetric transport.
We give an explicit solution for the full $N$-body
stationary distribution in cases where it factorizes, and briefly discuss
a physical application to a schematic turbulence modeling.

\begin{figure}[t]
\centering\includegraphics[width=7.5cm,clip]{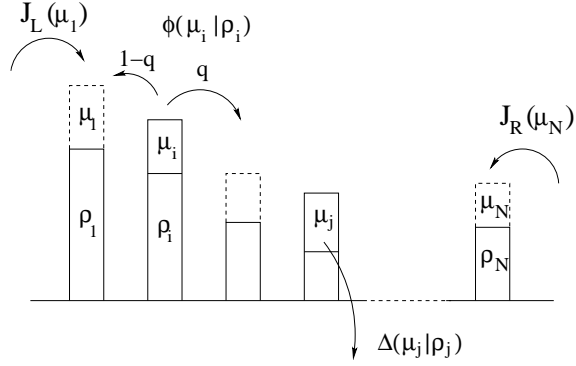}
\caption{\sl Schematic view of the model, with the three different mechanisms:
energy injection at boundary sites with rates $J_L(\mu_1)$ and $J_R(\mu_N)$,
transport from site $i$ to site $i+1$ and $i-1$ with rates
$q \phi_i(\mu_i|\rho_i)$ and $(1-q) \phi_i(\mu_i|\rho_i)$ respectively,
and dissipation at site $j$ with rate $\Delta_j(\mu_j|\rho_j)$.
On sites $i=1$ and $i=N$, energy may also be transferred to the reservoirs
--see text.
}
\label{fig-model}
\end{figure}

\section*{Definition of the model}

We consider a one-dimensional lattice with sites labelled from $i=1$ to $N$,
and introduce a local quantity $\rho_i$, that may be thought of as an
energy. The dynamics
is defined through the following local stochastic rules.
An amount of energy between $\mu$ and $\mu+d\mu$ can be
injected from a reservoir on the left boundary $i=1$ with a rate
(probability per unit time) $J_L(\mu) d\mu$, or injected from another
reservoir on the right boundary $i=N$ with a rate $J_R(\mu) d\mu$.
Transport proceeds by removing an amount of energy $\mu$
on site $i$ with a rate $\phi_i(\mu|\rho_i)d\mu$ depending on the energy
$\rho_i$ present on site $i$. The energy $\mu$ is then added either to
site $i+1$ (to the right reservoir if $i=N$) with probability $q$,
or to site $i-1$ (to the left reservoir if $i=1$)
with the complementary probability $p=1-q$.
Finally, dissipation is implemented by withdrawing an energy $\mu$ from
site $i$ with a rate $\Delta_i(\mu|\rho_i)d\mu$,
without adding it to another site.
A simple sketch of the model is shown on Fig.~\ref{fig-model}.

The statistical state of the model is described by the $N$-body probability
distribution $P(\rho_1,...,\rho_N,t)$.
Given the above stochastic rules, the evolution of this distribution
is governed by the following master equation:

\bea \nonumber
\frac{\partial P}{\partial t} =
q \sum_{j=1}^{N-1} \int_0^{\rho_{j+1}} d\mu\, \phi_j(\mu|\rho_j+\mu)
P(...,\rho_j+\mu,\rho_{j+1}-\mu,...,t)\\
\nonumber
+ p \sum_{j=2}^{N} \int_0^{\rho_{j-1}} d\mu\, \phi_j(\mu|\rho_j+\mu)
P(...,\rho_{j-1}-\mu,\rho_j+\mu,...,t)\\
\nonumber
+ \sum_{j=1}^N \int_0^{\infty} d\mu\, \Delta_j(\mu|\rho_j+\mu)
P(...,\rho_j+\mu,...,t)\\
\nonumber
+ p\int_0^{\infty} d\mu\, \phi_1(\mu|\rho_1+\mu) P(\rho_1+\mu,...,t)\\
\nonumber
+ q \int_0^{\infty} d\mu\, \phi_N(\mu|\rho_N+\mu) P(...,\rho_N+\mu,t)\\
\nonumber
+ \int_0^{\rho_1} d\mu\, J_L(\mu) P(\rho_1-\mu,...,t)
+ \int_0^{\rho_N} d\mu\, J_R(\mu) P(...,\rho_N-\mu,t)\\
\nonumber
- \sum_{j=1}^N \int_0^{\rho_j} d\mu\,
[\phi_j(\mu|\rho_j)+\Delta_j(\mu|\rho_j)] P(\{\rho_i\},t)\\
- \int_0^\infty d\mu\, [J_L(\mu)+J_R(\mu)]  P(\{\rho_i\},t)
\eea
where the dots stand for all the variables $\rho_i$
that are not modified by $\mu$.

\section*{Factorized steady state}

We wish to study the stationary distribution
$P_{st}(\rho_1,...,\rho_N)$ in cases where it is factorized, and
thus make the following ansatz for $P_{st}(\rho_1,...,\rho_N)$:
\be
P_{st}(\rho_1,...,\rho_N) = \frac{1}{Z} \prod_{i=1}^N f_i(\rho_i)
\ee
where $Z$ is a normalisation factor.
Note that due to the presence of dissipation in the bulk, one expects
that the one-site distribution $f_i(\rho_i)$ is different from one site
to the other, even when the stochastic rules are site-independent,
i.e., $\phi_i(\mu|\rho_i)=\phi(\mu|\rho_i)$ and
$\Delta_i(\mu|\rho_i)=\Delta(\mu|\rho_i)$.
Introducing the above ansatz for $P_{st}(\rho_1,...,\rho_N)$ in the stationary
master equation, it is easy to check that a specific form for the transport
and dissipation functions $\phi_i(\mu|\rho_i)$ and $\Delta_i(\mu|\rho_i)$
allows for a decoupling of the different degrees of freedom, in the
sense that the stationary master equation can then be written as a sum
of terms involving at most one variable $\rho_i$. This specific form reads:
\be \label{phi-delta}
\phi_i(\mu|\rho_i) = \tilde{\phi}_i(\mu) \,
\frac{f_i(\rho_i-\mu)}{f_i(\rho_i)}, \quad
\Delta_i(\mu|\rho_i) = \tilde{\Delta}_i(\mu) \,
\frac{f_i(\rho_i-\mu)}{f_i(\rho_i)}
\ee
Under these assumptions, the stationary master equation can be rewritten as:
\bea \nonumber
\sum_{j=1}^N \int_0^{\rho_j} d\mu\, [\tilde{\phi}_j(\mu)+
\tilde{\Delta}_j(\mu)] \frac{f_j(\rho_j-\mu)}{f_j(\rho_j)}
+ \int_0^{\infty} d\mu\, [J_L(\mu)+J_R(\mu)]\\
\nonumber
= q \sum_{j=2}^N \int_0^{\rho_j} d\mu\, \tilde{\phi}_{j-1}(\mu) 
\frac{f_j(\rho_j-\mu)}{f_j(\rho_j)}
+ p \sum_{j=1}^{N-1} \int_0^{\rho_j} d\mu\, \tilde{\phi}_{j+1}(\mu)
\frac{f_j(\rho_j-\mu)}{f_j(\rho_j)}\\
\nonumber
+ \int_0^{\infty} d\mu\, \left( q\tilde{\phi}_N(\mu) +
p \tilde{\phi}_1(\mu) + \sum_{j=1}^N \tilde{\Delta}_j(\mu) \right)\\
+ \int_0^{\rho_1} d\mu\, J_L(\mu) \frac{f_1(\rho_1-\mu)}{f_1(\rho_1)}
+ \int_0^{\rho_N} d\mu\, J_R(\mu) \frac{f_N(\rho_N-\mu)}{f_N(\rho_N)}
\eea
and the decoupling property mentioned above appears explicitly.
As this equation is a sum of functions of independent variables, each of these
functions --i.e., the sum of terms depending on a given $\rho_i$--
has to be equal to a constant.
To determine the value of these constants, one can
first send all the $\rho_i's$ to zero, in which case all the integrals
over the interval $[0,\rho_i]$ are expected to vanish
\footnote{Considering an integral of the form
$I(\rho) = \int_0^{\rho} d\mu\, \psi(\mu)f(\rho-\mu)/f(\rho)$ with
$\psi(\mu) \sim \mu^{\alpha-1}$ and $f(\mu) \sim \mu^{\beta-1}$ for
$\mu \to 0$ ($\alpha, \beta>0$ to ensure the convergence of the integral),
it can be shown easily that $I(\rho) \sim \rho^{\alpha} \to 0$ when
$\rho \to 0$.}.
As a result, the terms that do not depend on $\rho_i$ sum up to zero:
\be
\int_0^{\infty} d\mu\, \left( q\tilde{\phi}_N(\mu) + p\tilde{\phi}_1(\mu)
+\sum_{j=1}^N \tilde{\Delta}_j(\mu) - J_L(\mu) - J_R(\mu) \right) = 0
\ee
In a similar way, sending to zero all the $\rho_i$'s but one leads to the
conclusion that all terms depending on a given $\rho_j$ also sum up to zero.
For $2\le j \le N-1$, one has:
\be
\int_0^{\rho_j}d\mu\, [q\tilde{\phi}_{j-1}(\mu)
+ p\tilde{\phi}_{j+1}(\mu) - \tilde{\phi}_j(\mu) -\tilde{\Delta}_j(\mu)]
\frac{f_j(\rho_j-\mu)}{f_j(\rho_j)} = 0
\ee
At the boundary sites $j=1$ and $j=N$, one finds:
\bea
\int_0^{\rho_1}d\mu\, [J_L(\mu)
+ p\tilde{\phi}_2(\mu) - \tilde{\phi}_1(\mu) -\tilde{\Delta}_1(\mu)]
\frac{f_1(\rho_1-\mu)}{f_1(\rho_1)} = 0\\
\int_0^{\rho_1}d\mu\, [J_R(\mu)
+ q\tilde{\phi}_{N-1}(\mu) - \tilde{\phi}_N(\mu) -\tilde{\Delta}_N(\mu)]
\frac{f_N(\rho_N-\mu)}{f_N(\rho_N)} = 0
\eea
A sufficient condition to solve this set of equations is to assume that each
of the integrands is equal to zero, leading to:
\bea
\label{eq1}
q\tilde{\phi}_N(\mu) + p\tilde{\phi}_1(\mu)
+\sum_{j=1}^N \tilde{\Delta}_j(\mu) - J_L(\mu) - J_R(\mu) = 0\\
\label{eq2}
q\tilde{\phi}_{j-1}(\mu) + p\tilde{\phi}_{j+1}(\mu)
-\tilde{\phi}_j(\mu) -\tilde{\Delta}_j(\mu) = 0 \quad (2\le j \le N-1)\\
\label{eq3}
J_L(\mu) + p\tilde{\phi}_2(\mu) - \tilde{\phi}_1(\mu)
-\tilde{\Delta}_1(\mu) = 0\\
\label{eq4}
J_R(\mu) + q\tilde{\phi}_{N-1}(\mu) - \tilde{\phi}_N(\mu)
-\tilde{\Delta}_N(\mu) = 0
\eea
Thus, as a consequence of the specific form (\ref{phi-delta}) chosen for
$\phi_j(\mu|\rho)$ and $\Delta_j(\mu|\rho)$,
it turns out that the function $f_j(\rho)$
has disappeared from the equations --although this result may be surprising
at first sight, it should be noticed that the mass transport model
introduced in \cite{Evans04b} has the same property.
As a result, any set of functions $f_j(\rho)$ may be a solution of the model,
provided that the dynamical rules are suitably chosen.
Eqs.~(\ref{eq1}), (\ref{eq2}), (\ref{eq3}) and (\ref{eq4})
should thus be considered
as compatibility conditions allowing for the existence of a factorized
distribution.
Let us also note that these four equations are not independent: it can
be checked easily that, for instance, Eq.~(\ref{eq1}) can be obtained from
the sum of Eqs.~(\ref{eq2}) --summed over $j$--, (\ref{eq3}) and (\ref{eq4}).
To solve these equations, we first define some extra functions
$\tilde{\phi}_0(\mu)$ and $\tilde{\phi}_{N+1}(\mu)$ on the boundaries as
--assuming $p$, $q \ne 0$:
\be \label{boundary}
J_L(\mu) \equiv q\, \tilde{\phi}_0(\mu), \qquad
J_R(\mu) \equiv p\, \tilde{\phi}_{N+1}(\mu)
\ee
which allows one to rewrite Eqs.~(\ref{eq3}) and (\ref{eq4}) on the same form
as Eq.~(\ref{eq2}) with $j=1$ and $j=N$ respectively.
Introducing an auxiliary function $\chi_j(\mu)$ through:
\be \label{def-chi}
\chi_j(\mu) \equiv \tilde{\phi}_{j+1}(\mu) - \tilde{\phi}_j(\mu),
\qquad 0 \le j \le N
\ee
Eqs.~(\ref{eq2}), (\ref{eq3}) and (\ref{eq4}) can be rewritten in a
concise form as:
\be \label{eq-laplace}
p \chi_j(\mu) - q \chi_{j-1}(\mu) = \tilde{\Delta}_j(\mu),
\qquad 1 \le j \le N
\ee
Setting $r \equiv p/q$, this last equation is easily integrated out as:
\be
\chi_j(\mu) = \tilde{K}(\mu) \, r^{N-j}
- \frac{1}{q} \sum_{i=j}^{N-1} r^{i-j} \tilde{\Delta}_{i+1}
\qquad 0 \le j \le N
\ee
where $\tilde{K}(\mu)$ is (up to now) an arbitrary function of $\mu$,
and with the convention that the sum is zero if $j=N$.
Using Eq.~(\ref{def-chi}) and assuming $q\ne 1/2$,
one can now solve for $\tilde{\phi}_j(\mu)$, yielding for $ 0 \le j \le N+1$
--the sum is zero if $j \ge N$:
\be \label{phi-tilde}
\tilde{\phi}_j(\mu) = K(\mu) \, r^{N-j}
+ \frac{1}{2q-1} \sum_{i=j+1}^N (1-r^{i-j}) \tilde{\Delta}_i + C(\mu)
\ee
with a new (arbitrary) function $C(\mu)$, and where $K(\mu)$ is related to
$\tilde{K}(\mu)$ through $K(\mu)=p\tilde{K}(\mu)/(2q-1)$.
Using the boundary conditions (\ref{boundary}), one can determine the
unknown functions $K(\mu)$ and $C(\mu)$:
\bea
K(\mu) &=& B\, r
\left( \frac{J_R(\mu)}{1-q} - \frac{J_L(\mu)}{q} + \frac{1}{2q-1}
\sum_{i=1}^N (1- r^i) \tilde{\Delta}_i(\mu) \right)\\
C(\mu) &=& \frac{B}{q} \left(J_L(\mu) - J_R(\mu)\, r^N -
\frac{q}{2q-1} \sum_{i=1}^N (1- r^i) \tilde{\Delta}_i(\mu) \right)
\label{Cmu}
\eea
with $B \equiv (1-r^{N+1})^{-1}$.
Note that the case $q=1$, that was temporary excluded from the calculation,
is recovered by taking the limit $q \to 1$.
A similar calculation can be performed in the case $q=1/2$ to give:
\be \label{phi-tilde2}
\tilde{\phi}_j(\mu) = 2\sum_{i=1}^j (j-i)\tilde{\Delta}_i(\mu) + j A(\mu) +
2J_L(\mu) \qquad 1 \le j \le N
\ee
with
\be
A(\mu) = \frac{2}{N+1} \left[ J_R(\mu)-J_L(\mu) - \sum_{i=1}^N (N+1-i)
\tilde{\Delta}_i(\mu) \right]
\ee
As mentioned in the above derivation, the form (\ref{phi-delta}) of the rates
$\phi_i(\mu|\rho_i)$ and $\Delta_i(\mu|\rho_i)$, as well as the conditions
given in Eqs.~(\ref{phi-tilde}) and (\ref{phi-tilde2})
are a priori only sufficient conditions
for the factorization of the distribution $P_{st}(\rho_1,...,\rho_N)$.
Yet, let us recall that for the general class of mass transport models
on a ring geometry studied in \cite{Evans04b,Zia04}, where dissipation is
absent, the form (\ref{phi-delta}) of the transport rate has been shown to be
a necessary and sufficient condition for factorization, in the case of
continuous time dynamics. So it may be
plausible that this form is also necessary in the present model.
Let us also note that an explicit test of the form (\ref{phi-delta})
of the rate functions has been proposed in \cite{Zia04}.
This test should also apply to the present situation so that, given two
functions $\phi_i(\mu|\rho_i)$ and $\Delta_i(\mu|\rho_i)$, one should be
able to check whether they are of the form (\ref{phi-delta}), and to
determine the corresponding functions $\tilde{\phi}_i(\mu)$,
$\tilde{\Delta}_i(\mu)$ and $f_i(\rho)$.

\section*{Site-independent rates}

A situation of physical interest is when the dynamical rules are
site-independent whereas, due to the injection and dissipation mechanisms,
the one-site steady-state distribution may depend on the site considered.
An example of such a situation has been given in \cite{Bertin}, in the
fully biased case $q=1$. In what follows, we wish to study the possibility
to find a factorized steady-state distribution with symmetric transport,
that is for $q=1/2$, and with site-independent rates. The transition rates
given in Eq.~(\ref{phi-delta}) are independent of the site if the
functions $\tilde{\phi}_j(\mu)$, $\tilde{\Delta}_j(\mu)$ and $f_j(\rho)$
satisfy
\be \label{site-indep}
\tilde{\phi}_j(\mu) = h_1(\mu)\, e^{-\lambda_j \mu}, \: \:
\tilde{\Delta}_j(\mu) = h_2(\mu)\, e^{-\lambda_j \mu}, \: \:
f_j(\rho) = g(\rho)\, e^{-\lambda_j \rho}
\ee
Besides, from Eq.~(\ref{eq-laplace}), the rates have to satisfy
\be \label{eq-laplace2}
\frac{1}{2} \tilde{\phi}_{j-1}(\mu) + \frac{1}{2} \tilde{\phi}_{j+1}(\mu)
- \tilde{\phi}_j(\mu) = \tilde{\Delta}_j(\mu),
\qquad 1 \le j \le N
\ee
Let us introduce a coordinate $x=j/L$, so that $0 \le x \le 1$.
For the sake of simplicity, we shall consider the limit of large system size,
and assume that the transition rates vary slowly as a function of the site
index, in the sense that there exists continuous functions
$\hat{\phi}(\mu,x)$ and $\hat{\Delta}(\mu,x)$ of the variable $x$, such that
\be
\tilde{\phi}_j(\mu) = \hat{\phi}(\mu,jL), \qquad
\tilde{\Delta}_j(\mu) = \frac{1}{L^2} \hat{\Delta}(\mu,jL).
\ee
The $1/L^2$ factor has been included so that Eq.~(\ref{eq-laplace2}
admits a consistent continuous limit.
From Eq.~(\ref{site-indep}), this means that $\lambda_j$ must be replaced
by a continuous function $\lambda(x)$.
Accordingly, $\hat{\phi}(\mu,x)$ and $\hat{\Delta}(\mu,x)$ satisfy the
following equation
\be
\frac{\partial^2 \hat{\phi}}{\partial x^2}(\mu,x) = 2\hat{\Delta}(\mu,x)
\ee
which may be rewritten as
\be
\mu \lambda'(x)^2 - \lambda''(x) = \frac{2h_2(\mu)}{\mu\, h_1(\mu)}
\ee
As the r.h.s.~of this last equation does not depend on $x$, the only
possibility is that $\lambda(x)$ is linear in $x$, so that the l.h.s.~is also
a constant. As a result, it turns out that there is no factorized solution
that would be symmetric in $x$ with respect to $x=1/2$. This suggests
that the present model, with $q=1/2$ and symmetric injection rates
$J_L(\mu)=J_R(\mu)$, does not admit a factorized steady state,
at least in the continuous limit considered here.
The above calculation, with $\lambda(x)$ linear in $x$, rather corresponds
to a cascade process, where energy is injected on one boundary only,
in a way essentially similar to the fully biased case \cite{Bertin}.

\section*{A simple physical application}

Coming back to the value $q=1$ (fully biased model), one can try to use an
inhomogeneous version of the model (i.e., with site-dependent rates)
in order to describe schematically a turbulent cascade --note that only
a homogeneous case was studied in \cite{Bertin}. 
To this aim, we assume that the lattice sites $n$ may be thought of as
successive wavenumbers $k_n = n k_1$, and that energy is injected
on site $n=1$ (corresponding to the largest wavelength) with
a rate $J_L(\mu)=J_0\, e^{-\beta \mu}$, whereas there is
no energy transfer starting from site $n=N$, that is $J_R(\mu)=0$ and
$\phi_N(\mu|\rho)=0$. Note that the energy flux entering the system,
which in steady state equals the total rate of energy dissipation,
is given by $\varepsilon=J_0/\beta^2$.
To proceed further, the properties of the transition rates
have to be specified.
For simplicity, we assume $f_n(\rho)$ to take an
exponential form, namely $f_n(\rho)=e^{-\alpha_n \rho}$.
A characteristic feature of the turbulence phenomena is that the energy
dissipation rate $\Omega_n$ on site $n$ is of the form
$\Omega_n = \nu k_n^2 \langle \rho_n \rangle$,
where $\nu$ is a viscosity.
We thus compute $\Omega_n$ in the present model, yielding:
\be \label{eq-In}
\Omega_n = \frac{1}{Z_n} \int_0^{\infty} d\rho f_n(\rho)
\int_0^{\rho} d\mu\, \mu \Delta_n(\mu|\rho) =
\int_0^{\infty} d\mu\, \mu \tilde{\Delta}_n(\mu)
\ee
with $Z_n \equiv \int_0^{\infty} d\rho f_n(\rho)$, and where the last equality
is obtained by permuting the integrals and using the form (\ref{phi-delta}).
A second constraint arises from Eqs.~(\ref{phi-tilde}) to (\ref{Cmu})
encoding the factorization condition for the probability distribution,
which reads:
\be
J_L(\mu) = \sum_{n=1}^N \tilde{\Delta}_n(\mu)
\ee
A simple form for $\tilde{\Delta}_n(\mu|\rho)$ that matches the above
constraints is
\be
\tilde{\Delta}_n(\mu) = \frac{\nu k_1^2 n^2}{\alpha_n} \beta^2 e^{-\beta \mu}
\ee
on condition that the following self-consistency equation is satisfied:
\be \label{eq-J0}
\varepsilon = \sum_{n=1}^N \frac{\nu k_1^2 n^2}{\alpha_n}
\ee
Then, once $\tilde{\Delta}_n(\mu)$ is known, one can determine
$\tilde{\phi}_n(\mu)$ through Eqs.~(\ref{phi-tilde}) to (\ref{Cmu}).

In the following, we take the infinite $N$ limit and consider the small
viscosity regime in which developed turbulence is usually observed.
To determine the energy spectrum, i.e., the values of $\alpha_n$,
we require that the energy flux $\varepsilon$ does not depend on the
viscosity in the limit $\nu \to 0$.
In other words, the $\alpha_n$'s must be such that the
sum in Eq.~(\ref{eq-J0}) is independent of $\nu$.
Assuming that $k_1$ is small and that $\alpha_n$ depends smoothly on $n$,
one can replace the above sum by an integral over the wavenumber $k=k_1 n$,
introducing the energy density $E(k)$ at wavenumber $k$ through
$E(k_n)=1/(k_1 \alpha_n)$:
\be
\sum_{n=1}^{\infty} \frac{\nu k_1^2 n^2}{\alpha_n} 
= \sum_{n=1}^{\infty} \nu k_n^2 k_1 E(k_n)
\approx \int_0^{\infty} dk \, \nu k^2 E(k)
\ee
Making the assumption that the energy $\rho_n$ has the dimension of a velocity
square, it is well-known from Kolmogorov's K41 theory \cite{Frisch} that
a dimensionally consistent solution for $E(k)$ that yields a non-vanishing
energy flux for small $\nu$ is given by
\be
E(k)=\varepsilon^{\frac{2}{3}} k^{-\frac{5}{3}} g(\eta k),
\qquad \eta \equiv \left( \frac{\nu^3}{\varepsilon} \right)^{\frac{1}{4}}
\ee
where $\eta$ is the dissipation scale, and $g(x)$ is a scaling function
with a finite limit when $x \to 0$ and
such that $\int_0^{\infty} dx\, x^{1/3} g(x)=1$.
So our model allows one to construct a stochastic model for turbulence
obeying the K41 phenomenology. Its main drawback is that due to the
factorization condition, the energy spectrum must be somehow included by hand
in the stochastic rules, whereas one may rather wish this spectrum
to arise as a solution of the model, by including in the microscopic rules
only some basic physical principles. In this respect, it may be of great
interest to be able to go beyond the factorized steady-state distribution,
and one might try to solve this model with a more general matrix product
ansatz.

\section*{Conclusion}

In the present note, we have introduced a class of exactly solvable
stochastic models, which include both transport and dissipation phenomena.
The factorization condition for the probability distribution, as well as
an explicit solution for the latter, have been derived.
Some specific models within the class may used to illustrate
the appearance of (generalised) Gumbel distributions in complex systems
\cite{Bertin}, or to model, in a schematic way, the cascade phenomenon
in turbulent flows. More generally, one might hope
that this class of models could help to understand some fundamental issues
concerning the statistical behaviour of dissipative systems with a large
number of degrees of freedom.

\section*{Acknowledgements}
The author is grateful to P. Holdsworth and M. Clusel for discussions in an
early stage of this work, and to M. Droz for a critical reading of the
manuscript.


\section*{References}
\begin{thebibliography}{10}

\bibitem{Spitzer}
F. Spitzer, Adv. Math. {\bf 5}, 246 (1970).

\bibitem{Derrida}
B. Derrida, M.~R. Evans, V. Hakim and V. Pasquier, J. Phys. A: Math. Gen.
{\bf 26}, 1493 (1993).

\bibitem{Sandow}
S. Sandow, Phys. Rev. E {\bf 50}, 2660 (1994).

\bibitem{Liggett}
T.~M. Liggett, {\it Stochastic Models of Interacting Systems: Contact, Voter
and Exclusion Processes} (Springer, Berlin, 1999).

\bibitem{Schutz}
G.~M. Sch\"utz, Exactly solvable models for many-body systems far from
equilibrium, in {\it Phase Transitions and Critical Phenomena}, vol 19,
C. Domb and J. Lebowitz Eds (Academic, London, 2001).

\bibitem{Evans-Rev00}
M.~R. Evans, Braz. J. Phys. {\bf 30}, 42 (2000).

\bibitem{Evans-Rev05}
M.~R. Evans and T. Hanney, J. Phys. A: Math. Gen. {\bf 38}, R195 (2005).

\bibitem{Krug}
J. Krug and J. Garcia, J. Stat. Phys. {\bf 99}, 31 (2000).

\bibitem{Rajesh}
R. Rajesh and S.~N. Majumdar, J. Stat. Phys. {\bf 99}, 943 (2000).

\bibitem{Coppersmith}
S.~N. Coppersmith {\it et. al.}, Phys. Rev. E {\bf 53}, 4673 (1996).

\bibitem{Zielen}
F. Zielen and A. Schadschneider, J. Stat. Phys. {\bf 106}, 173 (2002).

\bibitem{Evans04b}
M.~R. Evans, S.~N. Majumdar and R.~K.~P. Zia, J. Phys. A: Math. Gen. {\bf 37},
L275 (2004).

\bibitem{Zia04}
R.~K.~P. Zia, M.~R. Evans and S.~N. Majumdar, J. Stat. Mech. : Theor. Exp.
L10001, (2004).

\bibitem{Evans02}
M.~R. Evans, Y. Kafri, E. Levine and D. Mukamel, J. Phys. A: Math. Gen.
{\bf 35}, L433 (2002).

\bibitem{Evans03}
M.~R. Evans, R. Juh\'asz and L. Santen, Phys. Rev. E {\bf 68}, 026117 (2003).

\bibitem{Evans04a}
M.~R. Evans, T. Hanney and Y. Kafri, Phys. Rev. E {\bf 70}, 066124 (2004).

\bibitem{Levine04}
E. Levine and R.~D. Willmann, J. Phys. A: Math. Gen. {\bf 37}, 3333 (2004).

\bibitem{Angel05}
A.~G. Angel, M.~R. Evans, E. Levine and D. Mukamel, preprint cond-mat/0503487.

\bibitem{Farago}
J. Farago, J. Stat. Phys. {\bf 118}, 373 (2005).

\bibitem{Bertin}
E. Bertin, Phys. Rev. Lett. {\bf 95}, 170601 (2005).

\bibitem{Frisch}
See e.g. U. Frisch, {\it Turbulence} (Cambridge University Press, 1995),
and references therein.

\end{thebibliography}
\end{document}